\documentclass[prl,twocolumn,twoside,showpacs,floatfix,superscriptaddress]{revtex4}

\usepackage{graphicx,color,rotating}
\usepackage{amsmath,amssymb}
\usepackage{dcolumn}
\usepackage{times,txfonts}
\usepackage{pifont}

\newcommand{\eq}[1]{\begin{equation}#1\end{equation}}
\newcommand{\eqmulti}[1]{\begin{equation}\begin{split}#1\end{split}\end{equation}}

\newcommand{\ket}[1]{\ensuremath{\,|{#1}\rangle}}

\newcommand{\matrixe}[3]{\ensuremath{\langle{#1}|\,{#2}\,|{#3}\rangle}}

\newcommand{\op}[1]{\ensuremath{#1}}
\newcommand{\adj}[1]{\ensuremath{{{#1}}^{\dag}}}

\newcommand{\elem}[2]{\ensuremath{{}^{#2}\text{#1}}}

\newcommand{\symboldiamond}[1][black]{{\color{#1}\hspace{-1pt}\footnotesize\begin{turn}{45} $\blacksquare$ \end{turn}}}
\newcommand{\symboltriangle}[1][black]{{\color{#1}$\blacktriangle$}}
\newcommand{\symbolbox}[1][black]{{\color{#1}$\blacksquare$}}
\newcommand{\symbolcircle}[1][black]{{\color{#1}\large$\bullet$}}

\definecolor{FGViolet}{rgb}{0.61,0.32,0.61}
\definecolor{FGDarkBlue}{rgb}{0,0,0.6}
\definecolor{FGBlue}{rgb}{0,0,0.8}
\definecolor{FGLightBlue}{rgb}{0.2, 0.6, 0.8}
\definecolor{FGGreen}{rgb}{0.2,0.7,0.2}
\definecolor{FGLightGreen}{rgb}{0.4,1,0.4}
\definecolor{FGYellow}{rgb}{1,0.95,0}
\definecolor{FGOrange}{rgb}{0.95,0.5,0.1}
\definecolor{FGRed}{rgb}{0.8,0,0}
\definecolor{FGWhite}{rgb}{1,1,1}
\definecolor{FGLightGray}{rgb}{0.8,0.8,0.8}
\definecolor{FGGray}{rgb}{0.5,0.5,0.5}
\definecolor{FGDarkGray}{rgb}{0.3,0.3,0.3}
\definecolor{FGBlack}{rgb}{0,0,0}

\begin{document}

\title{\emph{Ab Initio} Calculations of Medium-Mass Nuclei with Normal-Ordered Chiral NN+3N Interactions}

\author{Robert Roth}
\email{robert.roth@physik.tu-darmstadt.de}

\author{Sven Binder}
\author{Klaus Vobig}
\author{Angelo Calci}
\author{Joachim Langhammer}

\affiliation{Institut f\"ur Kernphysik, Technische Universit\"at Darmstadt, 64289 Darmstadt, Germany}

\author{Petr Navr\'atil}
\affiliation{TRIUMF, 4004 Wesbrook Mall, Vancouver, British Columbia, V6T 2A3, Canada}

\date{\today}

\begin{abstract}  

We study the use of truncated normal-ordered three-nucleon interactions in \emph{ab initio} nuclear structure calculations starting from chiral two- plus three-nucleon Hamiltonians evolved consistently with the similarity renormalization group (SRG). We present three key steps: (i) a rigorous benchmark of the normal-ordering approximation in the importance-truncated no-core shell model (IT-NCSM) for \elem{He}{4}, \elem{O}{16}, and \elem{Ca}{40}; (ii) a direct comparison of the IT-NCSM results with coupled-cluster calculations at the singles and doubles level (CCSD) for \elem{O}{16}; and (iii) first applications of SRG-evolved chiral NN+3N Hamiltonians in CCSD for the medium-mass nuclei \elem{O}{16,24} and \elem{Ca}{40,48}. We show that the normal-ordered two-body approximation works very well beyond the lightest isotopes and opens a path for \emph{ab initio} studies of medium-mass and heavy nuclei with chiral two- plus three-nucleon interactions. At the same time we highlight the predictive power of chiral Hamiltonians.

\end{abstract}

\pacs{21.60.De, 21.30.-x, 05.10.Cc, 21.45.Ff}

\maketitle

Two decades of experience with \emph{ab initio} nuclear structure calculations, including the pioneering work with the Green's Function Monte Carlo (GFMC) approach \cite{PuPa97}, have shown that three-nucleon (3N) interactions play an important role for understanding the structure of nuclei systematically from first principles. Recent advances on nuclear interactions derived within chiral effective field theory (EFT) put additional emphasis on the consistent inclusion of realistic 3N interactions in nuclear structure calculations \cite{MaEn11,BeEp08,BeEp11,EpNo02}. However, for most large-scale many-body approaches, such as the no-core shell model (NCSM) \cite{NaGu07} or the coupled-cluster (CC) method \cite{HaPa07}, the full inclusion of 3N interactions increases the computational cost by orders of magnitude and often renders calculations impossible that are routinely performed with two-nucleon (NN) Hamiltonians. In order to resolve this dilemma---the need for 3N interactions versus the tremendous computational cost---one might resort to approximate schemes for including the 3N interaction, particularly when aiming at a controlled approximation rather than an exact solution of the many-body problem. Along these lines, density-dependent NN interactions are being used for nuclear matter and finite nuclei to simulate the effects of 3N interactions in a computationally simple approximation (for recent applications see Refs.~\cite{LoBe11,HoKa10,HeSc10}). A related but more rigorous and improvable scheme to derive an approximate lower-rank form of the 3N interaction is based on a normal-ordered Hamiltonian with respect to a nucleus-specific reference state.

In this letter we present rigorous benchmarks and systematic applications of the normal-ordering approximation for chiral 3N interactions. We use the importance-truncated NCSM (IT-NCSM)  \cite{Roth09,RoNa07} for the \emph{ab initio} solution of the many-body problem for \elem{He}{4}, \elem{O}{16}, and \elem{Ca}{40} and compare ground-state energies and expectation values obtained with the full and the truncated normal-ordered 3N interaction. We then apply the normal-ordered 3N interaction in CC calculations for the ground-states of medium-mass nuclei up to \elem{Ca}{48} and discuss the implications for nuclear structure predictions with chiral NN+3N Hamiltonians.


\paragraph{Normal-Ordering Approximation.}

Transforming the many-body Hamiltonian into a normal-ordered form with respect to a nontrivial vacuum state is a standard technique in quantum many-body physics. Using creation and annihilation operators, $\op{a}^{\dag}_{\nu}$ and $\op{a}_{\nu}$, for a single-particle basis $\ket{\nu}$ defined with respect to a trivial zero-body vacuum state, the operator of a 3N interaction formally reads
\eq{ \label{eq:vacuumNO}
  \op{V}_{3N}
  = \frac{1}{36} \sum_{\substack{\nu_1\nu_2\nu_3\\ \mu_1\mu_2\mu_3}}  
    V^{\nu_1 \nu_2 \nu_3}_{\mu_1 \mu_2 \mu_3}\;
    \op{A}^{\nu_1 \nu_2 \nu_3}_{\mu_1 \mu_2 \mu_3} \;,
}
where $V^{\nu_1 \nu_2 \nu_3}_{\mu_1 \mu_2 \mu_3}=\matrixe{\nu_1\nu_2\nu_3}{\op{V}_{3N}}{\mu_1\mu_2\mu_3}$ are antisymmetrized matrix elements, and the operator $\op{A}^{\nu_1 \nu_2 \dots}_{\mu_1 \mu_2 \dots} = \adj{\op{a}}_{\nu_1} \adj{\op{a}}_{\nu_2} \cdots \op{a}_{\mu_2} \op{a}_{\mu_1}$ is a short-hand for a normal-ordered product of creation and annihilation operators  with respect to the trivial vacuum (vacuum normal-ordering). 

Instead of the trivial vacuum, we can choose a reference state $\ket{\Phi_{\text{ref}}}$ being a Slater-determinant of $A$ single-particle states and reinterpret the creation and annihilation operators as particle or hole creation and annihilation operators with respect to this new vacuum. We have to rearrange the creation and annihilation operators in the particle-hole picture to establish normal-ordering with respect to $\ket{\Phi_{\text{ref}}}$ (reference normal-ordering). Using Wick's theorem and $\widetilde{\op{A}}^{\nu_1 \nu_2 \dots}_{\mu_1 \mu_2 \dots}$ as a short-hand for the reference normal-ordered product, we obtain for the 3N interaction
\eqmulti{ \label{eq:referenceNO}
  \op{V}_{3N}
  &= W
    + \sum_{\substack{\nu_1\\ \mu_1}} 
      W^{\nu_1}_{\mu_1}\; \widetilde{\op{A}}^{\nu_1}_{\mu_1}
    + \frac{1}{4} \sum_{\substack{\nu_1\nu_2\\ \mu_1\mu_2}} 
      W^{\nu_1\nu_2}_{\mu_1\mu_2}\; \widetilde{\op{A}}^{\nu_1\nu_2}_{\mu_1\mu_2} \\
  &+ \frac{1}{36} \sum_{\substack{\nu_1\nu_2\nu_3\\ \mu_1\mu_2\mu_3}}  
      W^{\nu_1\nu_2\nu_3}_{\mu_1\mu_2\mu_3}\; \widetilde{\op{A}}^{\nu_1\nu_2\nu_3}_{\mu_1\mu_2\mu_3} \;,
} 
with matrix elements 
$W = \frac{1}{6}\sum_{\alpha_1 \alpha_2 \alpha_3} V^{\alpha_1 \alpha_2 \alpha_3}_{\alpha_1 \alpha_2 \alpha_3}$ 
for the zero-body (0B) term,
$W^{\nu_1}_{\mu_1}  = \frac{1}{2}\sum_{\alpha_2 \alpha_3}  V^{\nu_1 \alpha_2 \alpha_3}_{\mu_1 \alpha_2 \alpha_3}$ 
for the one-body (1B) term, 
$W^{\nu_1\nu_2}_{\mu_1\mu_2} = \sum_{\alpha_3}  V^{\nu_1 \nu_2 \alpha_3}_{\mu_1 \mu_2 \alpha_3}$
for the two-body (2B) term, and
$W^{\nu_1\nu_2\nu_3}_{\mu_1\mu_2\mu_3} = V^{\nu_1 \nu_2 \nu_3}_{\mu_1 \mu_2 \mu_3}$ 
for the residual three-body (3B) term, where $\alpha_i$ labels the occupied single-particle states in $\ket{\Phi_{\text{ref}}}$. 

The interesting aspect of the reference state as compared to the trivial vacuum is that it contains information about the specific many-body system under consideration---the reference state provides a first approximation to, e.g., the ground-state of a closed-shell nucleus. Based on this information, contributions of the 3N interaction are demoted to lower-particle ranks.
This is the basis of the normal-ordered $n$-body (NO$n$B) approximation.

\paragraph{Benchmark of the Normal-Ordering Approximation.}

To quantify how well such a truncation works, we perform \emph{ab initio} calculations for the ground states of closed-shell nuclei in the IT-NCSM. The IT-NCSM allows us to include the exact 3N interaction just as well as any NO$n$B approximation. The underlying Hamiltonian contains the chiral NN interaction at N${}^3$LO of Ref.~\cite{EnMa03} and the local chiral 3N interaction at N${}^2$LO of Ref.~\cite{Navr07}. The low-energy constants $c_D$ and $c_E$ are taken from a fit to the ground-state energy and the $\beta$-decay half-life of $A=3$ systems \cite{GaQu09}. This initial Hamiltonian is transformed through a similarity renormalization group (SRG) evolution at the two- and three-body level to enhance the convergence behavior of the many-body calculation \cite{RoLa11,RoNe10,JuNa09}. The SRG evolution represents a continuous unitary transformation parametrized by a flow-parameter $\alpha$, with $\alpha=0$ corresponding to the initial Hamiltonian. We will mainly consider two types of SRG-evolved Hamiltonians: The NN+3N-full Hamiltonian starts with the initial chiral NN+3N Hamiltonian and retains all terms up to the 3N level in the SRG-evolution, the NN+3N-induced Hamiltonian omits the chiral 3N interaction from the initial Hamiltonian, but keeps all induced 3N terms throughout the evolution. The 3N terms in both Hamiltonians have quite different characteristics, which makes them useful for benchmarking the normal-ordering approximation. In addition we will employ a range of values of the flow parameter $\alpha$ to generate an even larger set of test cases.

\begin{figure}
\includegraphics[width=1\columnwidth]{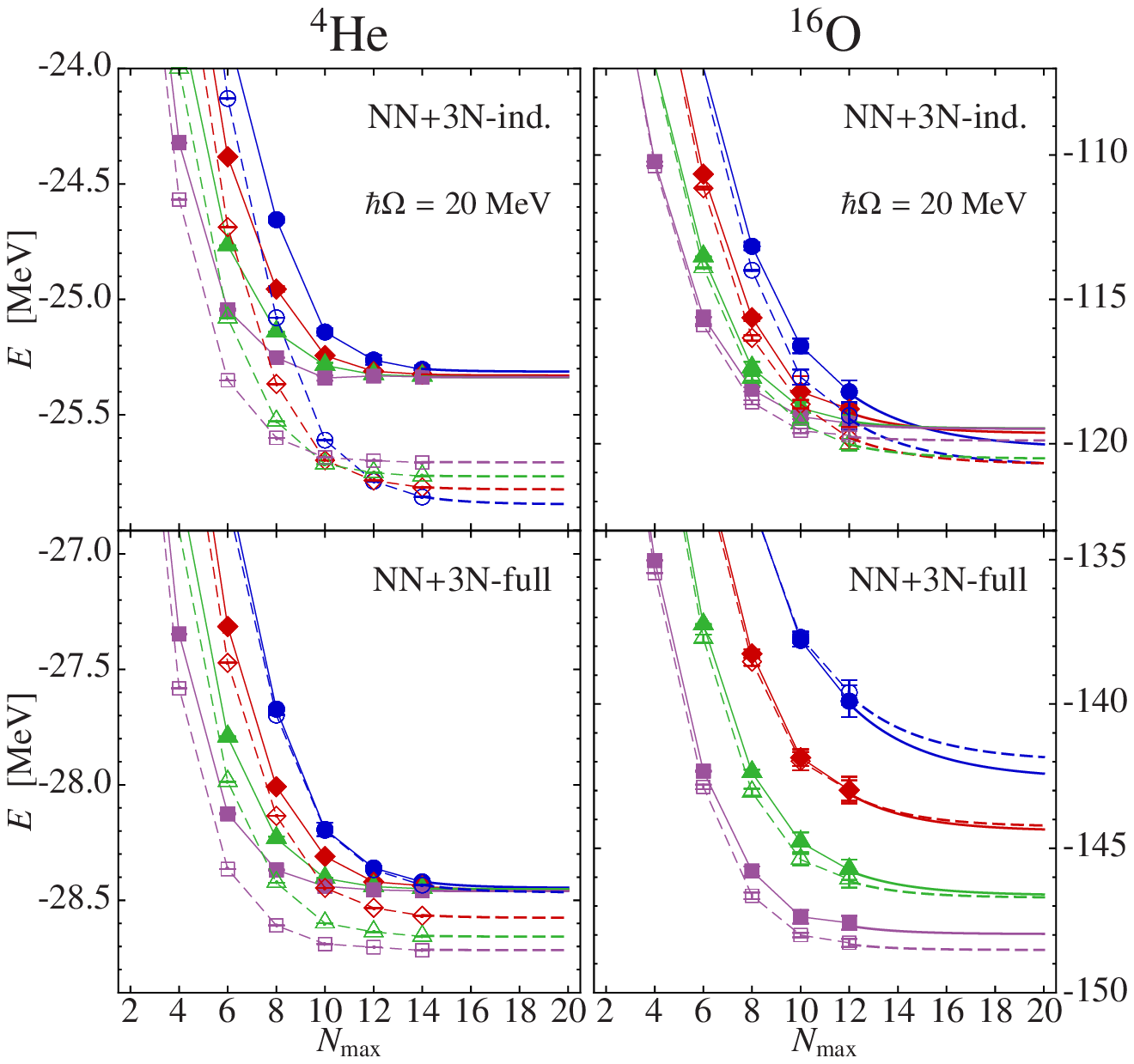}
\caption{(color online) IT-NCSM ground-state energies for \elem{He}{4} and \elem{O}{16} as function of $N_{\max}$ for the NN+3N-induced and the NN+3N-full Hamiltonians for a range of flow parameters: $\alpha=0.04\,\text{fm}^4$ (\symbolcircle[FGBlue]), $0.05\,\text{fm}^4$ (\symboldiamond[FGRed]), $0.0625\,\text{fm}^4$ (\symboltriangle[FGGreen]), $0.08\,\text{fm}^4$ (\symbolbox[FGViolet]). Solid symbols correspond to the exact 3N interaction, open symbols to the NO2B approximation. Error bars indicate the uncertainties of the threshold extrapolations of the IT-NCSM. Data points are connected by straight lines to guide the eye, beyond the largest $N_{\max}$ an exponential extrapolation fitted to the last four data points is shown. }
\label{fig:itncsm_He4O16}
\end{figure}

We start with a direct comparison of IT-NCSM calculations for the ground-state energies of \elem{He}{4} and \elem{O}{16} using either the exact 3N interaction or the NO2B approximation. The reference state is always the $0\hbar\Omega$ Slater-determinant composed of harmonic-oscillator single-particle states. The IT-NCSM energies as function of the model-space truncation parameter $N_{\max}$ are presented in Fig.~\ref{fig:itncsm_He4O16}. The comparison of the converged values of the ground-state energy obtained with the NO2B approximation and with the exact 3N interaction reveals a multifaceted picture. The largest relative deviation at the level of 2\% is observed for \elem{He}{4} with the NN+3N-induced interaction. The NO2B approximation leads to a overbinding of about 0.4 MeV for the softest Hamiltonian ($\alpha=0.08\,\text{fm}^4$) and 0.6 MeV for the hardest Hamiltonian ($\alpha=0.04\,\text{fm}^4$). This systematics is completely reversed when using the NN+3N-full Hamiltonian, here the softest interaction produces the largest deviation of about 0.3 MeV, whereas for the hardest interaction the NO2B approximation coincides with the exact 3N calculations. For \elem{O}{16} we observe deviations below 1\% for the ground-state energy, again with a nontrivial dependence on the Hamiltonian. For NN+3N-induced the NO2B approximation consistently overestimates the binding energy by about 1 MeV, for NN+3N-full the NO2B approximation overbinds by less than 1 MeV for $\alpha=0.08\,\text{fm}^4$ and underbinds by less than 1 MeV for $\alpha=0.04\,\text{fm}^4$. 

\begin{figure}
\includegraphics[width=1\columnwidth]{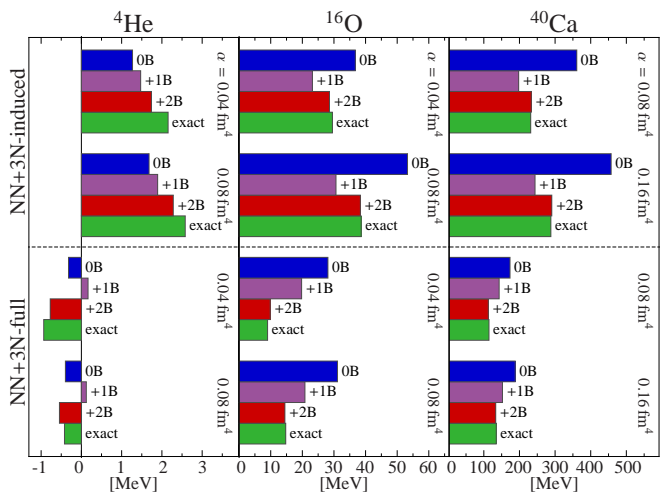}
\caption{(color online) Anatomy of the NO$n$B approximation of the ground-state energies of \elem{He}{4}, \elem{O}{16}, and \elem{Ca}{40}. The bar charts show the expectation values of the 3N interaction computed  at different levels of the normal-ordering approximation, i.e., NO0B, NO1B, NO2B, and exact 3N. We employ the NN+3N-induced and NN+3N-full Hamiltonians, each with two values of $\alpha$ (see labels). We use the eigenstates obtained for the exact 3N interaction in $N_{\max}=10$ for \elem{He}{4} and \elem{O}{16} and $N_{\max}=8$ for \elem{Ca}{40}, all at $\hbar\Omega=20\,\text{MeV}$.}
\label{fig:itncsm_noanatomy}
\end{figure}

For a comprehensive picture of its anatomy, we analyze the expectation values of the 3N interaction at different levels of the NO$n$B approximation using IT-NCSM eigenstates obtained with the exact 3N interaction for \elem{He}{4}, \elem{O}{16}, and \elem{Ca}{40} for fixed $N_{\max}$. Figure \ref{fig:itncsm_noanatomy} summarizes these expectation values of the 3N interaction for a set of NN+3N-induced and NN+3N-full Hamiltonians. For \elem{O}{16} and \elem{Ca}{40} a similar pattern emerges: The NO2B approximation does reproduce the expectation value of the exact 3N interaction very well, both for the NN+3N-induced and the NN+3N-full Hamiltonian. The pattern observed for the sequence of NO$n$B approximations is different for both types of Hamiltonians. For NN+3N-induced the 1B and 2B contributions of the normal-ordered Hamiltonian have opposite sign, with the 1B contribution being significantly larger, whereas for the NN+3N-full Hamiltonian the 1B and 2B contributions are both attractive and of similar size. In all cases the 0B contribution is the largest and overestimates the exact 3N expectation value. For \elem{He}{4} the pattern is different. The 0B term does not provide the largest contribution and underestimates the 3N expectation value. The signs and relative sizes of the 1B and 2B terms again depend on the Hamiltonian and the NO2B approximation still shows a sizable deviation from the exact 3N expectation value---consistent with the discussion of Fig.~\ref{fig:itncsm_He4O16}.

This case study shows that there is no universal pattern and no hierarchy in the individual NO$n$B contributions. The size of the individual terms and also the deviation of the NO2B approximation from the exact 3N result depends on the Hamiltonian, the nucleus, and the oscillator frequency. Nonetheless, the 3N expectation values in Fig.~\ref{fig:itncsm_noanatomy} and the ground-state energies in Fig.~\ref{fig:itncsm_He4O16} demonstrate that the NO2B approximation works very well beyond the lightest nuclei.

\paragraph{Application in Coupled-Cluster Theory.}

\begin{figure}
\includegraphics[width=0.85\columnwidth]{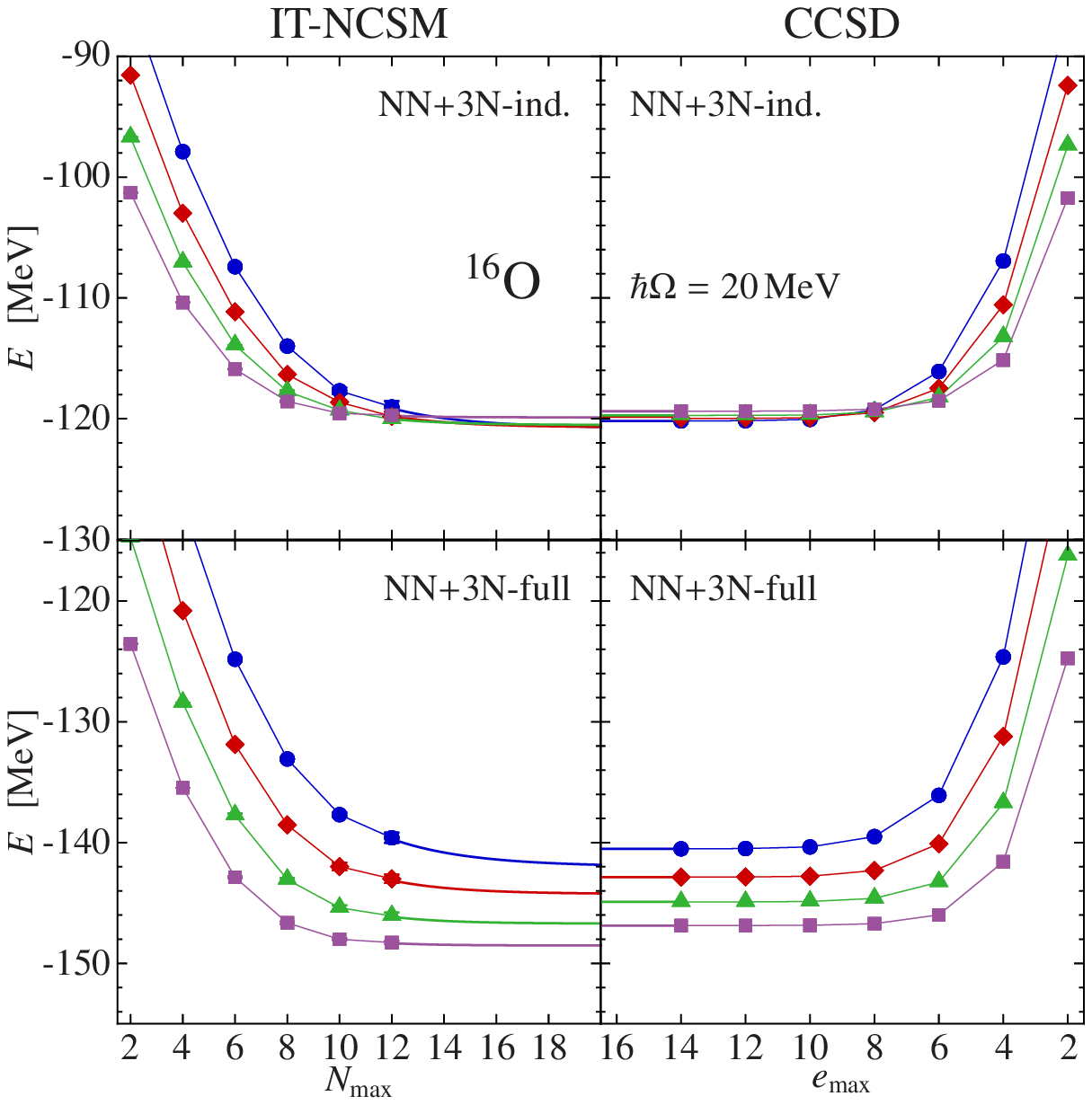}
\caption{(color online) Comparison of the ground-state energies of \elem{O}{16} obtained in IT-NCSM and CCSD including 3N interactions at the NO2B level for the NN+3N-induced and the NN+3N-full Hamiltonians with $\alpha=0.04\,\text{fm}^4$ (\symbolcircle[FGBlue]), $0.05\,\text{fm}^4$ (\symboldiamond[FGRed]), $0.0625\,\text{fm}^4$ (\symboltriangle[FGGreen]), and $0.08\,\text{fm}^4$ (\symbolbox[FGViolet]).}
\label{fig:itncsm_ccsd}
\end{figure}

After validating the NO2B approximation, we are now applying it in ground-state calculations for heavier closed-shell nuclei in the framework of the coupled-cluster method. Coupled-cluster theory is a natural framework, since normal-ordering of the Hamiltonian with respect to a reference state is inherent to the formulation of the approach. We have developed an efficient coupled-cluster code using the $J$-coupled scheme discussed in Ref.~\cite{HaPa10}, which enables us to go to very large model spaces. We limit ourselves to singles and doubles excitations (CCSD), which has been shown to be a good approximation for soft SRG-evolved interactions \cite{HaPa10}. An additional approximation present in the CCSD calculations for technical reasons is a truncation of the 3N matrix elements entering the NO2B to harmonic-oscillator principal quantum numbers $e_1+e_2+e_3 \leq E_{3\max} = 14$. Unlike the $N_{\max}$ truncation of the IT-NCSM, the $e_{\max}$ truncation of CCSD does not imply this 3B energy cut so that we miss some 3N matrix elements for $e_{\max}>6$. 

In a first step, we confront the CCSD results for \elem{O}{16} with the previous IT-NCSM results, both using the NO2B approximation. Figure \ref{fig:itncsm_ccsd} shows the convergence of the ground-state energies in both methods using the NN+3N-induced and NN+3N-full Hamiltonian. We observe a very good agreement of the converged ground-state energies, with the IT-NCSM giving 1 to 2 MeV more binding. This difference is consistent with the contributions expected from triples corrections and the missing 3N matrix elements with $E_{3\max}>14$. The latter point has been confirmed by comparing to lower $E_{3\max}$ cuts. Altogether, the CCSD calculations for \elem{O}{16} with soft SRG-evolved NN+3N Hamiltonians in NO2B approximation provide a ground-state energy within 1 to 2\% of the IT-NCSM results with the exact 3N interaction.

\begin{figure}
\includegraphics[width=1\columnwidth]{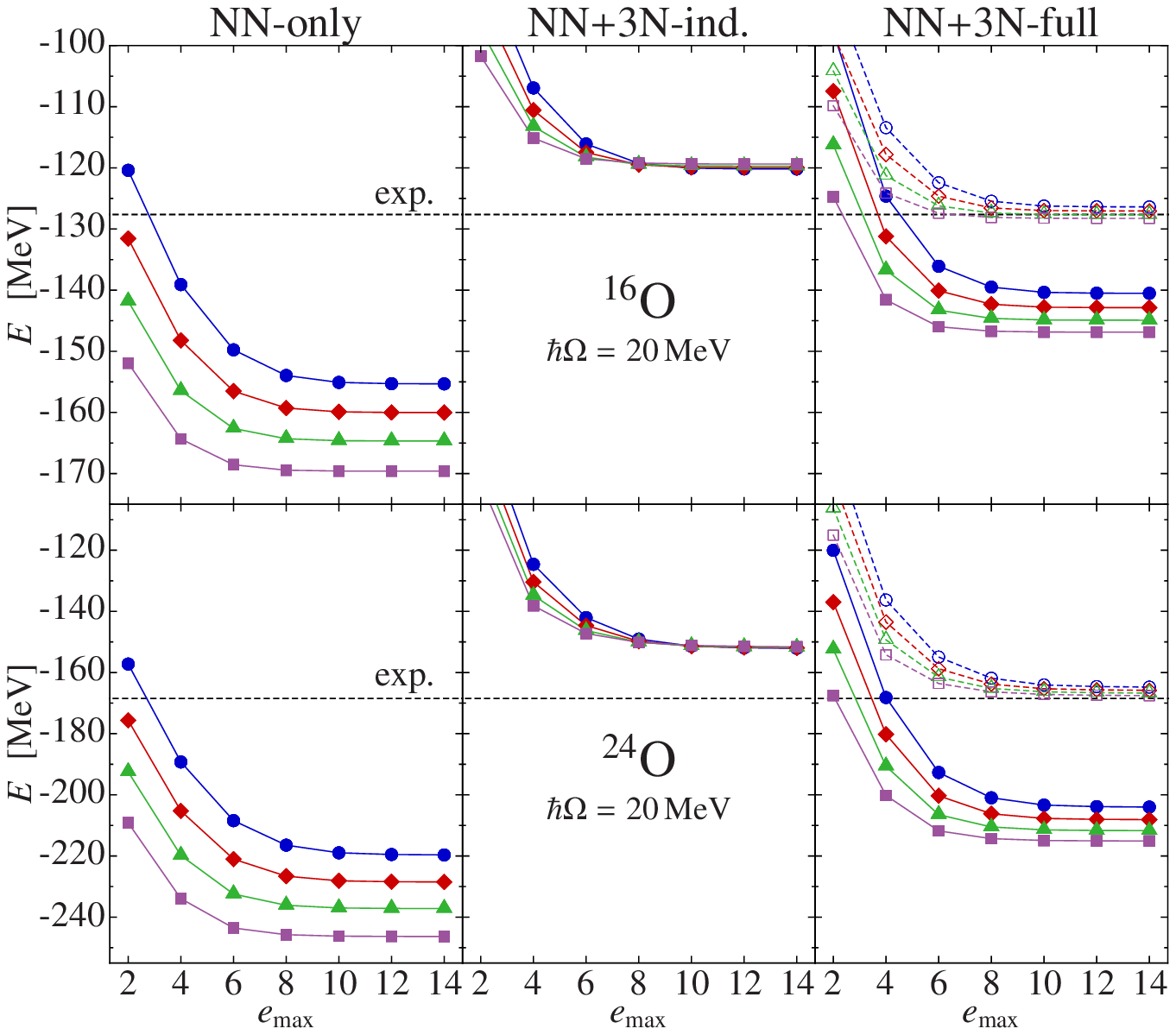}
\caption{(color online) CCSD ground-state energies for \elem{O}{16} and \elem{O}{24} as function of $e_{\max}$ for the three types of Hamiltonians (see column headings) using the NO2B approximation for a range of flow parameters: $\alpha=0.04\,\text{fm}^4$ (\symbolcircle[FGBlue]), $0.05\,\text{fm}^4$ (\symboldiamond[FGRed]), $0.0625\,\text{fm}^4$ (\symboltriangle[FGGreen]), and $0.08\,\text{fm}^4$ (\symbolbox[FGViolet]). The filled symbols for the NN+3N-full Hamiltonian are for the standard chiral 3N interaction with cutoff 500 MeV, the open symbols for a modified 3N interaction with cutoff 400 MeV (see text).}
\label{fig:ccsd_O16O24}
\end{figure}
\begin{figure}
\includegraphics[width=1\columnwidth]{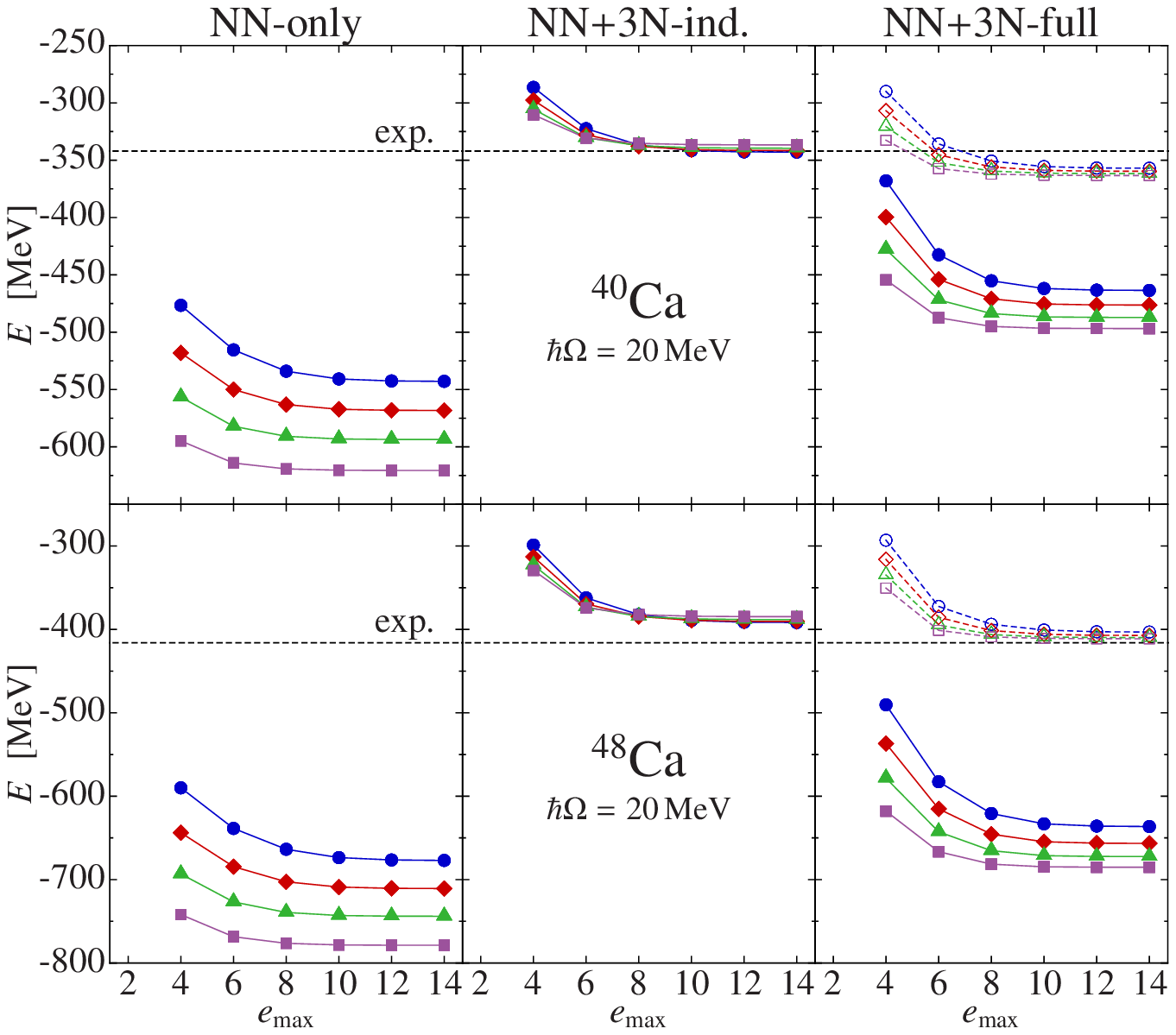}
\caption{(color online) Same as Fig.~\ref{fig:ccsd_O16O24} for \elem{Ca}{40} and \elem{Ca}{48}.}
\label{fig:ccsd_Ca40Ca48}
\end{figure}

Using CCSD with the NO2B approximation we can now study the systematics of ground-state energies with SRG-evolved chiral NN+3N Hamiltonians beyond \elem{O}{16}. Following the analysis of Ref.~\cite{RoLa11} we discuss the $\alpha$-dependence observed with the NN-only, the NN+3N-induced, and the NN+3N-full Hamiltonians for \elem{O}{16} and \elem{O}{24}, shown in Fig. \ref{fig:ccsd_O16O24}, and for \elem{Ca}{40} and \elem{Ca}{48}, shown in Fig.~\ref{fig:ccsd_Ca40Ca48}. For all nuclei we observe the same pattern: The NN-only Hamiltonian exhibits strong $\alpha$-dependence of the converged ground-state energies hinting at induced 3N interactions. Their inclusion at the NN+3N-induced level eliminates the $\alpha$-dependence, thus demonstrating that induced 4N contributions originating from the initial NN are irrelevant. The converged energies, therefore, correspond to the solutions for the initial chiral NN interaction. We obtain $-120.2\,(+0.8)$ MeV for \elem{O}{16} ground-state energy, $-152.1\,(+0.5)$ MeV for \elem{O}{24}, $-343\,(+6)$ MeV for \elem{Ca}{40}, and $-392\,(+7)$ MeV for \elem{Ca}{48} using the NN+3N-induced Hamiltonian at $\alpha=0.04\,\text{fm}^4$ for $e_{\max}=14$. The numbers in parenthesis give the change when going to $\alpha=0.08\,\text{fm}^4$ as a measure for the residual $\alpha$-dependence. These results are in very good agreement with the CC results reported in Refs.~\cite{HaPa09b,HaPa10} for the bare chiral NN interaction. 

When including the initial 3N interaction, i.e., when using the NN+3N-full Hamiltonian, the $\alpha$-dependence reemerges, indicating that 4N terms induced by the initial 3N interaction become sizable. These CCSD results confirm the findings of Ref.~\cite{RoLa11} and extend the systematics to heavier nuclei. 

In addition to the standard chiral 3N interaction \cite{GaQu09} with cutoff momentum of $500\,\text{MeV}$, we also employ a chiral 3N interaction with a modified cutoff of $400\,\text{MeV}$ and $c_E=0.098$ refitted to reproduce the \elem{He}{4} binding energy. We keep the value $c_D=-0.2$ as in the standard 3N interaction. Effectively the lower cutoff reduces the strength of the two-pion terms of the 3N interaction and limits them to lower momenta. As a result the $\alpha$-dependence and thus the induced 4N contributions are reduced significantly. This allows for a quantitative comparison of the NN+3N-full predictions with experimental binding energies. We obtain ground-state energies of $-126.4\,(-1.9)$ MeV for \elem{O}{16}, $-164.8\,(-2.8)$ MeV for \elem{O}{24}, $-357\,(-6)$ MeV for \elem{Ca}{40}, and $-403\,(-8)$ MeV for \elem{Ca}{48} using $\alpha=0.04\,\text{fm}^4$ with the change when going to $\alpha=0.08\,\text{fm}^4$ given in parenthesis. The agreement with experiment is remarkable. For \elem{O}{16} and \elem{O}{24} the predictions based on the chiral NN+3N Hamiltonian are consistent with the experimental binding energies. Even for \elem{Ca}{40} and \elem{Ca}{48} the agreement with experiment is surprisingly good, given the fact that no information beyond \elem{He}{4} was used to fix the Hamiltonian. This is evidence that this chiral NN+3N Hamiltonian, although not fully consistent regarding the chiral order and the cutoff choice in NN and 3N terms, contains all the relevant physics for nuclear structure predictions over a large mass range.

\paragraph{Conclusions.}

We have demonstrated that the NO2B approximation allows for accurate nuclear structure calculations using SRG-evolved chiral NN+3N Hamiltonians in cases, where the inclusion of the exact 3N interaction is computationally too demanding. Therefore, it provides a valuable tool to exploit the full physics potential of chiral Hamiltonians. In this context we have shown that a chiral 3N interaction with reduced cutoff can yield binding energy systematics consistent with experiment---ongoing investigations of the spectroscopy of p- and sd-shell nuclei confirm the quality and universality of this Hamiltonian.  A generalization of the normal-ordering approximation to open-shell systems and multi-determinantal reference states is subject of present research.

\paragraph{Acknowledgments.}

Numerical calculations have been performed at the J\"ulich Supercomputing Centre and at LOEWE-CSC. Supported by the DFG through contract SFB 634, the Helmholtz International Center for FAIR (HIC for FAIR), and the BMBF (06DA9040I). P. N. acknowledges support from NSERC grant No. 401945-2011.


\end{document}